\documentclass[fdp,a4paper,fleqn%
]{w-art}

\newcommand{\id}{\relax{\rm 1\kern-.28em 1}}
\newcommand{\R}{\mathbb{R}}
\newcommand{\C}{\mathbb{C}}

\newcommand{\g}{\mathfrak{G}}

\newcommand{\rspan}{\mathrm{span}}
\newcommand{\ri}{\mathrm{i}}
\newcommand{\re}{\mathrm{e}}

\newcommand{\rGL}{\mathrm{GL}}
\newcommand{\rSU}{\mathrm{SU}}
\newcommand{\rSL}{\mathrm{SL}}
\newcommand{\rSO}{\mathrm{SO}}
\newcommand{\M}{\mathrm{M}}
\newcommand{\rM}{\mathrm{M}}
\newcommand{\rE}{\mathrm{E}}
\newcommand{\rcof}{\mathrm{cof}}

\newcommand{\hv}{ \hat{t}}

\newcommand{\cA}{\mathcal{A}}

\newcommand{\cO}{\mathcal{O}}

\usepackage{times,w-thm}
\theoremstyle{plain}

\theoremstyle{definition}

\usepackage[]{graphicx}
\usepackage{amsmath, amsthm}
\usepackage{amscd}
\usepackage{amssymb}
\usepackage{array}

\begin{document}
\pagespan{1}{}



\title[Deformation of Minkowski space]{Quadratic deformation of Minkowski space}


\author[D. Cervantes]{Dalia Cervantes\inst{1,}\footnote{E-mail:~\textsf{daliac@nucleares.unam.mx}}}%

\address[\inst{1}]{Instituto de Ciencias Nucleares, Universidad Nacional
Aut\'{o}noma de M\'{e}xico,
Circuito Exterior M\'{e}xico D.F. 04510, M\'{e}xico.}
\author[R. Cervantes]{Rita Fioresi\inst{2,}\footnote{E-mail:~\textsf{fioresi@dm.unibo.it}}}
\address[\inst{2}]{Dipartimento di Matematica,  Universit\`{a} di
Bologna. Piazza di Porta S. Donato, 5. 40126 Bologna. Italy.}
\author[M. A. Lled\'{o}]{Mar\'{\i}a A. Lled\'{o}\inst{3,}\footnote{Corresponding author\quad E-mail:~\textsf{maria.lledo@ific.uv.es},
            Phone: +00\,34\,9635\,43252,
            Fax: +00\,34\,9635\,43381}}
\address[\inst{3}]{Departament de F\'{\i}sica Te\`{o}rica,
Universitat de Val\`{e}ncia and Institut de F\'{\i}sica Corpuscular  (CSIC-UVEG). C/Dr.
Moliner, 50, E-46100 Burjassot (Val\`{e}ncia), Spain.}
%
\author[F. A. Nadal]{Felip A. Nadal\inst{3,}\footnote{E-mail:~\textsf{Felip.Nadal@ific.uv.es}}}
\begin{abstract}
  We present a deformation of the Minkowski space as embedded into the conformal space (in the formalism of twistors) based in the quantum versions of the corresponding kinematic groups. We compute explicitly the star product, whose Poisson bracket is quadratic. We show that the star product although defined on the polynomials can be extended differentiably. Finally we compute the Eucliden and Minkowskian real forms of the deformation.
\end{abstract}
\maketitle                   






\section{The Grassmannian and the Minkowski space}
We consider the  Grassmannian $G(2,4)$, the set of two-planes inside $\mathbb{C}^4$.
A plane $\pi\in G(2,4)$ is given by two linearly independent vectors or by any two linear combinations of them that are independent, so
$$\pi=\rspan(a, b)=\rspan(a', b'),\qquad (a', b')=(a,b) h,\qquad h\in \rGL(2, \mathbb{C}).$$
There is a transitive action of $\rGL(4,\C)$ (or $\rSL(4,\mathbb{C}))$ ) on $G(2,4)$.
$$ g\pi=\rspan \{ga, gb\},\qquad g\in \rGL(4,\mathbb{C}).$$
If we consider a particular point $\pi_0$
$$ \pi_0=\rspan\begin{pmatrix}1&0\\0&1\\\hline 0&0\\0&0\end{pmatrix},\qquad \hbox{with stability group}$$
$$P_0=\left\{\begin{pmatrix}L&M \\
0&R\end{pmatrix}\in \rSL(4,\mathbb{C})\right\},\qquad  \hbox{and} \qquad G(2,4)=\rSL(4,\mathbb{C})/P_0.$$

We notice that the conformal group of space time,
$\rSO(2,4)$, has spin group $\rSU(2,2)$. Its complexification, $\rSO(6,\C)$, has spin group $\rSL(4,\C)$.

How to extract the Minkowski space from $G(2,4)$? Notice that since the two vectors are independent,
$$\pi=\begin{pmatrix}a_1&b_1\\a_2&b_2\\
a_3&b_3\\a_4&b_4\end{pmatrix},$$ at least one of the $2\times 2$ determinants in this matrix is $\neq 0$. The space is covered by the atlas
$$U_{ij}=\left\{(a,b)\in \C^4\times \C^4\quad/\quad a_ib_j-b_ia_j\neq
0\right\},\; i<j, \quad i,j=1,\dots 4. $$
$U_{12}$ is the {\it big cell}, and using the $\rGL(2,\C)$ freedom, a plane in $U_{12}$ can be represented by $$\pi=\begin{pmatrix}1&0\\0&1\\t_{31}&t_{32}\\t_{41}&t_{42}\end{pmatrix}=\begin{pmatrix}\id\\t\end{pmatrix},$$ with the entries  of $t$ totally arbitrary. So $U_{12}\approx \C^4$, and it is a good candidate for the Minkowski space.

What about the group action? $U_{12}$ is left invariant by the {\it lower parabolic} subgroup of $\rSL(4,\mathbb{C})$,
$$P_l=\left\{\begin{pmatrix}x&0\\Tx&y\end{pmatrix},\quad /\quad  \det x\cdot\det y=1\right\},$$ and it acts on $t$ as
$$t\mapsto ytx^{-1}+T.$$ The group is $\rSL(2,\mathbb{C})\times \rSL(2,\mathbb{C})\times
\C^\times\ltimes T^4,$ so it is the Poincar\'{e} group where instead of the Lorentz group we have put its double cover.

$t$ belongs to the {\it twistor space} associated to spacetime.
Using the Pauli matrices, we can revert to the spacetime notation and obtain the standard action of the Poincar\'{e} group on Minkowski spacetime
$$t=x^\mu\sigma_\mu =\begin{pmatrix}
x^{0}+x^{3} & x^{1}-\ri x^{2} \\
x^{1}+ix^{2} & x^{0}-x^{3}
\end{pmatrix}.$$
Also the spacetime metric has an interpretation in the twistor formalism, $$\det t=(x^0)^2-(x^1)^2-(x^2)^2-(x^3)^2.$$

\section{Algebraic approach}

Quantization of spacetime means to deform the commutative algebra of functions (can be polynomials or smooth functions) to a non commutative algebra. Other properties that we want to consider in the quantum setting have to be first defined in the algebraic formalism and then 'quantized'. This is the case of the group actions.
The respective algebras are
$$\cO(\rSL(4,\C))=\C[g_{AB}]/ (\det g-1),\qquad A,B=1,\dots , 4.$$
$$\cO(P_l)=\C[x_{ij},  y_{a b},  T_{ai}]/( \det x\cdot \det y -1), \qquad   i,j=1,2,\quad  a,b=3,4.$$
$$\cO(\rM)=\C[t_{31},t_{32},t_{41},t_{42}].$$
The group law is expressed in terms of a {\it coproduct}
$$\begin{CD}\cO(\rSL(4,\C))@>\Delta_c>> \cO(\rSL(4,\C))\otimes \cO(\rSL(4,\C))\\g_{AB}@>>>\sum_Cg_{AC}\otimes g_{CB},\end{CD}\qquad  C=1,\dots,4,$$with the property
$$\mu_G\circ(\Delta_c f) (g_1, g_2)=f(g_1g_2),\qquad f\in \cO(\rGL(4,\C)).$$

The action on the Minkowski space is a {\it coaction}
$$\begin{CD} \cO(\rM)@>\tilde \Delta>>\cO(P_l)\otimes\cO(\rM)\\ t_{ai}@>>> y_{ab}S( x)_{ji}\otimes t_{bj}+ T_{ai}\otimes 1.\end{CD}$$

\section{The quantum Minkowski space}

In Refs. \cite{fi,cfl1} one substitutes the group $\rSL(4,\C)$ by $\rSL_q(4,\C)$in the twistor construction. All the scheme of coaction and big cell can be repeated in the quantum case, which gives a quantization for the Minkowski space as a big cell inside a quantum conformal space (a quantum Grassmannian).

We just state the  result: The quantum Minkowski space is a quantum matrix algebra with the rows interchanged. The correspondence $M_q(2)\rightarrow \cO_q(\M)$ is given in terms of the respective generators:
$$\begin{pmatrix}\hat{a}_{11}& \hat{a}_{12}\\ \hat{a}_{21}& \hat{a}_{22}\end{pmatrix}\rightleftarrows\begin{pmatrix}  \hv_{32}& \hv_{31}\\ \hv_{42}& \hv_{41}\end{pmatrix}.$$

This means that the commutation relations among the quantum generators are the following
\begin{align*}
&\hv_{42} \hv_{41}  =  q^{-1} \hv_{41} \hv_{42}, \qquad
&\hv_{31} \hv_{41}  =  q^{-1} \hv_{41} \hv_{31}, \qquad
& \hv_{32} \hv_{41}  =   \hv_{41} \hv_{32} + (q^{-1} -q ) \hv_{42} \hv_{31},\nonumber\\
& \hv_{31} \hv_{42}  =  \hv_{42} \hv_{31}, \qquad
& \hv_{32} \hv_{42}  =  q^{-1} \hv_{42} \hv_{32}, \qquad
& \hv_{32} \hv_{31}  =  q^{-1} \hv_{31} \hv_{32}.
\end{align*}
 What happened to the groups? They have become {\it quantum groups} with a non commutative product and a coproduct that is {\it the same} than the one  we had before. This means that the group law has not changed, nor the coaction on the Grassmannian and the Minkowski space. The only change is that all these varieties have become non commutative. It is a remarkable property of matrix quantum groups that the coproduct is compatible with both, the commutative product and the non commutative one.

In the quantum version  $\cO_q(P_l)$, the sets of generators $x$, $y$ and $ T$ are separately isomorphic to $2\times 2$ matrix algebras, but while $x$ and $y$ commute among them, $T$ does not commute with the rest of the generators.

\section{Algebraic star product}

A quantum matrix algebra is an algebra over $\C_q=\C[q,q^{-1}]$, where $q$ is a parameter. Moreover, as a module over $\C_q$, it is a free module, which means that it has a basis. It is well known that there is at least one {\it ordering } among the generators such that the standard monomials associated to this ordering are a basis of the quantum matrix algebra. (This is a non trivial property).

The ordering is the following
$$\hat t_{41}<\hat t_{42}<\hat t_{31}<\hat t_{32},$$
and then there is an isomorphism ({\it ordering rule} or {\it quantization map}) between $\cO_q(\M)$ and $\cO(\M)[q,q^{-1}]$:

$$\begin{CD}\cO(\M)[q,q^{-1}]@> Q_\M>>\cO_q(\M)\\t_{41}^a t_{42}^b t_{31}^c  t_{32}^d @>>> \hv_{41}^a \hv_{42}^b  \hv_{31}^c  \hv_{32}^d\end{CD}.$$

With the quantization map we can pull back the non commutative product to $\cO(\M)[q,q^{-1}]$. This defines a {\it star product},

$$f\star g=Q_\rM^{-1}\bigl(Q_{\rM}(f)Q_\rM(g)\bigr),\qquad f,g\in \cO(\rM)[q,q^{-1}],$$
which can be computed {\it explicitly}
\begin{align*}
&(t_{41}^a t_{42}^b t_{31}^c t_{32}^d )\star (t_{41}^m t_{42}^n t_{31}^p t_{32}^r)  =  q^{-mc-mb-nd-dp} t_{41}^{a+m} t_{42}^{b+n} t_{31}^{c+p} t_{32}^{d+r}  \\\ &+ \sum_{k = 1}^{\mu=min(d,m)} q^{-(m-k)c -(m-k)b - n(d-k)-p(d-k)}  F_k(q,d,m)  t_{41}^{a+m-k} t_{42}^{b+k+n} t_{31}^{c+k+p} t_{32}^{d-k+r}
\end{align*}

$F_k(q,d,m)$ are numerical factors defined recursively. We recover the semiclassical interpretation of the algebra being an algebra of functions, but with a star product.

\section{Differential star product}

The previous formula for the star product is nice and compact, but can only be computed on polynomials. Can we extend it to smooth functions? Not obvious. We prove that there exists a (unique) differential star product that coincides with the one given above on polynomials.

Change of the parameter: $q=\re ^h$. We expand in powers of $h$ so we obtain a star product of the form
$$f\star g= fg+ \sum_{j=1}^\infty h^j C_j(f, g),$$
with $$
f = t_{41}^a t_{42}^b t_{31}^c t_{32}^d, \qquad  g = t_{41}^m t_{42}^n t_{31}^p t_{32}^r.$$

At each order, we have contributions from each of the terms with different $k$
$$C_n(f,g)= \sum_{k=0}^{\mu=min(d,m)} C_n^{(k)}(f,g). $$
We want to write $C_n$ as a bidifferential operator. But this is not trivial because all the dependence in the exponents should cancel.
\begin{align*}
C_1  =& C_1^{(0)}+C_1^{(1)}=  - ( t_{41} t_{31} \partial_{31} \otimes \partial_{41}  + t_{42} t_{41} \partial_{42} \otimes \partial_{41}  +\\ & t_{32} t_{42} \partial_{32} \otimes \partial_{42} +     t_{32} t_{31} \partial_{32} \otimes \partial_{31}  + 2 t_{42} t_{31} \partial_{32} \otimes \partial_{41}  ),
\end{align*}
 Antisymmetrizing and changing variables we obtain the Poisson bracket
\begin{align*}
&\{f,g \}= \ri\Big((x^{0})^{2}-(x^{3})^{2})(\partial_{1}f\partial_{2}g -\partial_{1}g\partial_{2}f)  + x^{0}x^{1}(\partial_{0}f\partial_{2}g -\partial_{0}g\partial_{2}f)\\&
-x^{0}x^{2}(\partial_{0}f\partial_{1}g
-\partial_{0}g\partial_{1}f)-x^{1}x^{3}(\partial_{2}f\partial_{3}g -\partial_{2}g\partial_{3}f)
+x^{2}x^{3}(\partial_{1}f\partial_{3}g -\partial_{1}g\partial_{3}f)\Big)
\end{align*}

Notice that if the $x$'s are real, then the Poisson bracket is pure imaginary.   Also, it is quadratic.

We have computed explicitly up to the order $h^2$, but the expression is already too big to display it here.   We looked for an argument to show it at arbitrary order. This can be done by  careful inspection.
The proof that it is differential at each order is rather technical and we do not reproduce it here \cite{cfl2} But having the explicit formula for the polynomials is essential to apply the argument.

An example of the possible difficulty: suppose that we want to reproduce $x^{m-1}$ as the result of applying a differential operator to $x^m$. We have several choices,

$$x^{m-1}=\frac 1 x x^m,\qquad x^{m-1}=\frac 1 m\partial_x x^m,\qquad ... $$
But none of them is both, independent on the exponent $m$ and polynomial in the variable. So the right combination of coefficients should appear in order to cancel the factors that appear when differentiating. For example, if the result were $m x^{m-1}$, then we have a differential operator
$$mx^{m-1}=\partial_x x^m.$$

Since we have recovered the interpretation of 'functions' for the non commutative algebra, we can try to express the coaction as an action of this space of functions. Remember that, formally, for the generators the coaction is the same than in the commutative algebra. We just need to pull it back to the star product algebra.

We define the transformed variables (no translations are considered here)
$$
\tau_{ij} \equiv \mu_{G\times \M}  \circ \tilde\Delta_{\star}( t_{ij} ) = y_{ab} t_{bj}S(x_{ji});\qquad \hbox{so}
$$

$$
\mu_{G\times \M} \circ \Delta_{\star}( t_{41}^{ a} t_{42}^{b} t_{31}^{c} t_{32}^{d} )    = \tau_{41}^{\star a}  \star_{G\times \M} \tau_{42}^{\star b}  \star_{G\times \M}  \tau_{31}^{\star c} \star_{G\times \M} \tau_{32}^{\star d}.
$$   One just has to expand the star products in the right hand side. Up to order $h$ we have computed it in terms of a differential operator,
\begin{align*}
&  \frac{ 1}{2} D_1 ( \tau_{41}  , \tau_{41}  ) \partial_{\tau_{41} }^2   +      D_1 ( \tau_{41}  , \tau_{42}  )  \partial_{\tau_{41} } \partial_{\tau_{42} }  +    \frac{1}{2} D_1 ( \tau_{42}  , \tau_{42}  )  \partial_{\tau_{42} }^2  +    D_1 ( \tau_{42}  , \tau_{31}  ) \partial_{\tau_{42} } \partial_{\tau_{31} }  + \\&    \frac{1}{2} D_1 ( \tau_{31}  , \tau_{31}  ) \partial_{\tau_{31} }^2  +     D_1 ( \tau_{31}  , \tau_{32}  ) \partial_{\tau_{31} } \partial_{\tau_{32} } +   \frac{1}{2} D_1 ( \tau_{32}  , \tau_{32}  ) \partial_{\tau_{32} }^2  +     D_1 ( \tau_{41}  , \tau_{31}  ) \partial_{\tau_{41} } \partial_{\tau_{31} }  + \\&     D_1 ( \tau_{41}  , \tau_{32}  ) \partial_{\tau_{41} } \partial_{\tau_{32} } +     D_1 ( \tau_{42}  , \tau_{32}  ) \partial_{\tau_{42} } \partial_{\tau_{32} }.
\end{align*}
and the coefficients are polynomials of order 6 in the variables $x, y, t$.

\section{Real forms: Euclidean and Minkowski quantum spaces}

Let $\cA$ be a commutative algebra over $\C$. An involution $\iota$ of $\cA$ is an antilinear map satisfying, for $f,g\in \cA$ and $\alpha,\beta \in\C$
\begin{align*}
&\iota(\alpha f+\beta g)=\alpha^*\iota f +\beta^* \iota g,\qquad &\hbox{(antilinearity)}\\
&\iota(fg)=\iota (f)\iota(g),\qquad &\hbox{(automorphism)}\\
&\iota\circ\iota =\id.\qquad &
\end{align*}
Let us consider the set of fixed points of $\iota$,
$$\cA^\iota =\{f\in \cA\;/\; \iota(f)=f\}.$$
It is easy to see that this is a real algebra whose complexification is $\cA$. $\cA^\iota $ is a {\it real form} of $\cA$.

{\it Classical Minkowski space}:
$$\begin{pmatrix}\iota_\rM(t_{31})&\iota_\rM(t_{32})\\\iota_\rM(t_{41})&\iota_\rM(t_{42})\end{pmatrix}=
\begin{pmatrix}t_{31}&t_{41}\\t_{32}&t_{42}\end{pmatrix},\qquad \iota_\rM(t)=t^T.$$
The combinations
\begin{align*}&x^{0}=\frac 12(t_{31}+t_{42}),\qquad &x^{1}=\frac 12(t_{32}+t_{41}),\\&x^{2}=\frac 1 {2\ri} (t_{41}-t_{32}),&x^{3}=\frac 12(t_{31}-t_{42}),\\\end{align*}
are fixed points of the involution.    One has
$$\cO(\M)^{\iota_\M}=\R[x^0, x^1, x^2, x^3].$$

{\it Classical Euclidean space}:
$$\begin{pmatrix}\iota_\rE(t_{31})&\iota_\rE(t_{32})\\\iota_\rE(t_{41})&\iota_\rE(t_{42})\end{pmatrix}=
\begin{pmatrix}t_{42}&-t_{41}\\-t_{32}&t_{31}\end{pmatrix},\qquad \iota_\rE(t)=\rcof(t).$$

The commbinations
\begin{align*}&z^{0}=\frac 12(t_{31}+t_{42}),\qquad &z^{1}=\frac \ri 2(t_{32}+t_{41}),\\&z^{2}=\frac 1 {2} (t_{41}-t_{32}),&z^{3}=\frac \ri 2(t_{31}-t_{42}),\\\end{align*}
are
fixed points of $\iota_\rE$,   and as before,
$$\cO(\rM)^{\iota_E}=\R[z^0, z^1, z^2, z^3 ].$$

Formally, the same expressions on the generators as in the classical case are involutions in the quantum algebra. A few things change:
\begin{itemize}
\item Checking with the commutation relations they are antiautomorphisms, this is
\begin{align*}&\iota_{\rM}(fg)=\iota_{\rM}(g)\iota_{\rM}(f),\\
&\iota_{\rE}(fg)=\iota_{\rE}(g)\iota_{\rE}(f).
\end{align*}
\item This discards the interpretation of the real form of the non commutative algebra as the set of
fixed points of the involution.
\item When pulling back to the star product algebra, the Poisson bracket is purely imaginary.
\end{itemize}

Finally one finds also the corresponding involutions in the group,

\begin{align*}
&\iota_{P_l,\rM}(x)=S(y)^T, \qquad &\iota_{P_l,\rM}(y)=S(x)^T,\qquad &\iota_{P_l,\rM}(T)=T^T;\\
&\iota_{P_l,\rE}(x)=S(x)^T, \qquad &\iota_{P_l,\rE}(y)=S(y)^T,\qquad &\iota_{P_l,\rE}(T)=\rcof(T),
\end{align*}
It is not difficult to realize that in the Minkowskian case the real form of the Lorentz group (corresponding to the generators $x$ and $y$) is $\rSL(2,\C)_\R$ and in the Euclidean case is $\rSU(2)\times \rSU(2)$.

\section*{Acknowledgments}

D. Cervantes wants to thank the Departament de F\'{\i}sica Te\`{o}rica,
Universitat de Val\`{e}ncia for the hospitality
during the elaboration of this work.

Felip A. Nadal wants to thank CSIC for a JAE-predoc grant.

This work has been supported in part by grants FIS2008-06078-C03-02 and FIS2011-29813-C02-02 of Ministerio de Ciencia e Innovaci\'{o}n (Spain) and ACOMP/2010/213 of Generalitat Valenciana.

\end{document}